\documentclass{article}

\usepackage{microtype}
\usepackage{graphicx}
\usepackage{subfigure}
\usepackage{booktabs} % for professional tables
\usepackage{amsmath,amssymb,amsfonts}

\usepackage{hyperref}

\usepackage[accepted]{icml2021}

\icmltitlerunning{Human Perception of Computer Vision Interpretability}

\begin{document}

\twocolumn[
\icmltitle{Quality Metrics for Transparent Machine Learning \\With and Without Humans In the Loop Are Not Correlated}




\begin{icmlauthorlist}
\icmlauthor{Felix Biessmann}{bht,ecdf}
\icmlauthor{Dionysius Refiano}{bht}
\end{icmlauthorlist}

\icmlaffiliation{bht}{Beuth University of Applied Sciences, Berlin}
\icmlaffiliation{ecdf}{Einstein Center Digital Future, Berlin}

\icmlcorrespondingauthor{Felix Biessmann}{felix.biessmann@beuth-hochschule.de}

\icmlkeywords{Explainable AI, XAI, Interpretable Machine Learning, }

\vskip 0.3in
]



\printAffiliationsAndNotice{\icmlEqualContribution} % otherwise use the standard text.

\begin{abstract}
The field explainable artificial intelligence (XAI) has brought about an arsenal of methods to render Machine Learning (ML) predictions more interpretable. But how useful explanations provided by transparent ML methods are for humans remains difficult to assess. 
Here we investigate the quality of interpretable computer vision algorithms using techniques from psychophysics. In crowdsourced annotation tasks we study the impact of different interpretability approaches on annotation accuracy and task time. We compare these quality metrics with classical XAI, automated quality metrics. Our results demonstrate that psychophysical experiments allow for robust quality assessment of transparency in machine learning. Interestingly the quality metrics computed without humans in the loop did not provide a consistent ranking of interpretability methods nor were they representative for how useful an explanation was for humans. 
These findings highlight the potential of methods from classical psychophysics for modern machine learning applications. We hope that our results provide convincing arguments for evaluating interpretability in its natural habitat, human-ML interaction, if the goal is to obtain an authentic assessment of interpretability.
\end{abstract}

%!TEX root = manuscript.tex

\section{Introduction}
\label{sec:intro}

In recent years complex machine learning (ML) models, many based on deep learning, have achieved surprising results in computer vision, natural language processing and many other domains. These models are difficult to interpret, which inspired many researchers to investigate ways to render ML models {\em interpretable}~\cite{Kim2015,lipton2016mythos,doshi2017towards,Herman2017ThePA}. There are many motivations for interpretable ML methods. Domain experts, data scientists or data engineers that control proper functioning of an ML pipeline need to be be able to access the rules learned by a ML system in an intuitive manner in order to quickly spot the root causes of errors. More generally the main motivation for research on transparent ML is that intuitive human understanding of ML predictions can is a prerequisite for a healthy trust relationship between humans and assistive ML systems. In particular transparency is argued to prevent algorithm aversion as well as algorithmic bias. Algorithm aversion refers to cases when humans do not trust ML systems, even when they know that the model predictions are more accurate than those of a human \cite{Dietvorst2015}, algorithmic bias are cases of ethnical or gender biases in ML predictions \cite{Hajian2016}. In the following we will also use the term algorithmic bias to refer to cases of too much trust into an ML prediction, for instance when a human interacting with assistive ML technology blindly follows its predictions. The usual narrative is that explanations of ML decisions can increase human trust in them \cite{Sinha2002, lime}. 
%
%In recent years complex discriminative ML models, many based on deep learning, have achieved surprising results in computer vision, natural language processing and many other domains. These complex models were difficult to interpret, which inspired many researchers to investigate ways to render ML models {\em interpretable}.

A central problem with interpretability methods is that they are difficult to compare and evaluate. Most of the research  compares methods using either proxy measures, that do not directly relate to interpretability by humans, as e.g. \cite{Samek2017}, or qualitative measures that render comparisons of results across studies difficult \cite{Strumbel2010}. In this work we propose to use psychophysical methods to quantify and compare the quality of interpretability methods.
We follow the ideas of \cite{schmidt2019quantifying} and base our approach on the assumption that the definition of interpretability is inherently tied to a human observer. Good interpretability methods should allow human observers to intuitively understand a ML prediction. Intuitive understanding of the rules learned by a ML system is reflected in how accurately and how fast humans make decisions when assisted with a transparent ML prediction. These two variables can be easily measured in psychophysical experiments that study the interaction between humans and ML systems. 

The motivation for this work is twofold: For one this work aims at complementing previous work on measuring the quality of interpretability methods by establishing a quantitative measure of interpretability in the domain of computer vision that captures aspects of human cognition. Ultimately this will help practitioners to choose the right interpretability method for a given use case and researchers to devise novel objectives for better interpretability methods. Secondly the goal of this study is to validate to what extent existing approaches for measuring interpretability without humans in the loop reflect the interpretability metrics we measure in psychophysical experiments. 

In the following we shortly highlight some of the related work and then describe an image annotation task, emotion recognition, as well as the ML model, the transparency approaches used and the experimental design for quantitatively evaluating interpretability with humans in the loop (HIL) and with no humans in the loop (NHIL). We compare the different interpretability approaches with respect to the HIL and NHIL metrics and analyze their relationship, in particular whether cheaper and more scalable machine based NHIL transparency metrics reflect the most relevant but more expensive HIL transparency metrics. We conclude with highlighting the implications of our results for practitioners that build systems with human-ML interaction or transparent ML. 
%\input{related_work}
%!TEX root = psychophysics-XAI-4page.tex

\section{Experiments}
\label{sec:experiments}
%In the following we describe the annotation task and the experimental setup.
%
\paragraph{Annotation Task}
The annotation task was emotional expression classification on images. We used the extended Cohn-Kanade image data set \cite{Lucey2010} which contains images for the classes, anger, contempt, disgust, fear, happiness, sadness, surprise. 
%The class distribution can be seen in \autoref{tab:eckcd}. 
We reduced the data to a binary classification task in which annotators had to classify emotional expressions of {\em anger} and {\em happiness}. 
%Some sample images are shown in \autoref{fig:example}. 
The annotators had the option of not providing an annotation in case they did not recognize the emotional expression. 

\paragraph{Machine Learning Model}
We used a computer vision model from an open source python toolkit that achieves state of the art performance on emotional expression prediction from images \cite{emopy}. %We used the default model without any modifications. 
%The precision, recall, and F1 scores on the data set used in our experiments are shown in \autoref{tab:emopyscore}. Note that these predictive performances are not perfect, but they can be considered competitive with the state of the art for this particular classification task. We also emphasize that we here are focussing on interpretability methods, not the underlying ML model. The quality of the machine learning model was the same for all interpretability methods. 

%\begin{table}[]
%\begin{tabular}{llllllll}
%\toprule
%emotion & precision & recall & f1-score & support  \\
%\midrule
%anger & 0.48 & 0.47 & 0.47 & 45 \\
%happiness & 0.66 & 0.67 & 0.66 & 69 \\
%\midrule
%avg / total & 0.59 & 0.59 & 0.59 & 114 \\ 
%\bottomrule
%\end{tabular}
%\caption{Held-out per label precision/recall/f1 scores of EmoPy used for comparing ML interpretability methods on the CK+ dataset}
%\label{tab:emopyscore}
%\end{table}

%
\begin{figure}
\includegraphics[width=\columnwidth]{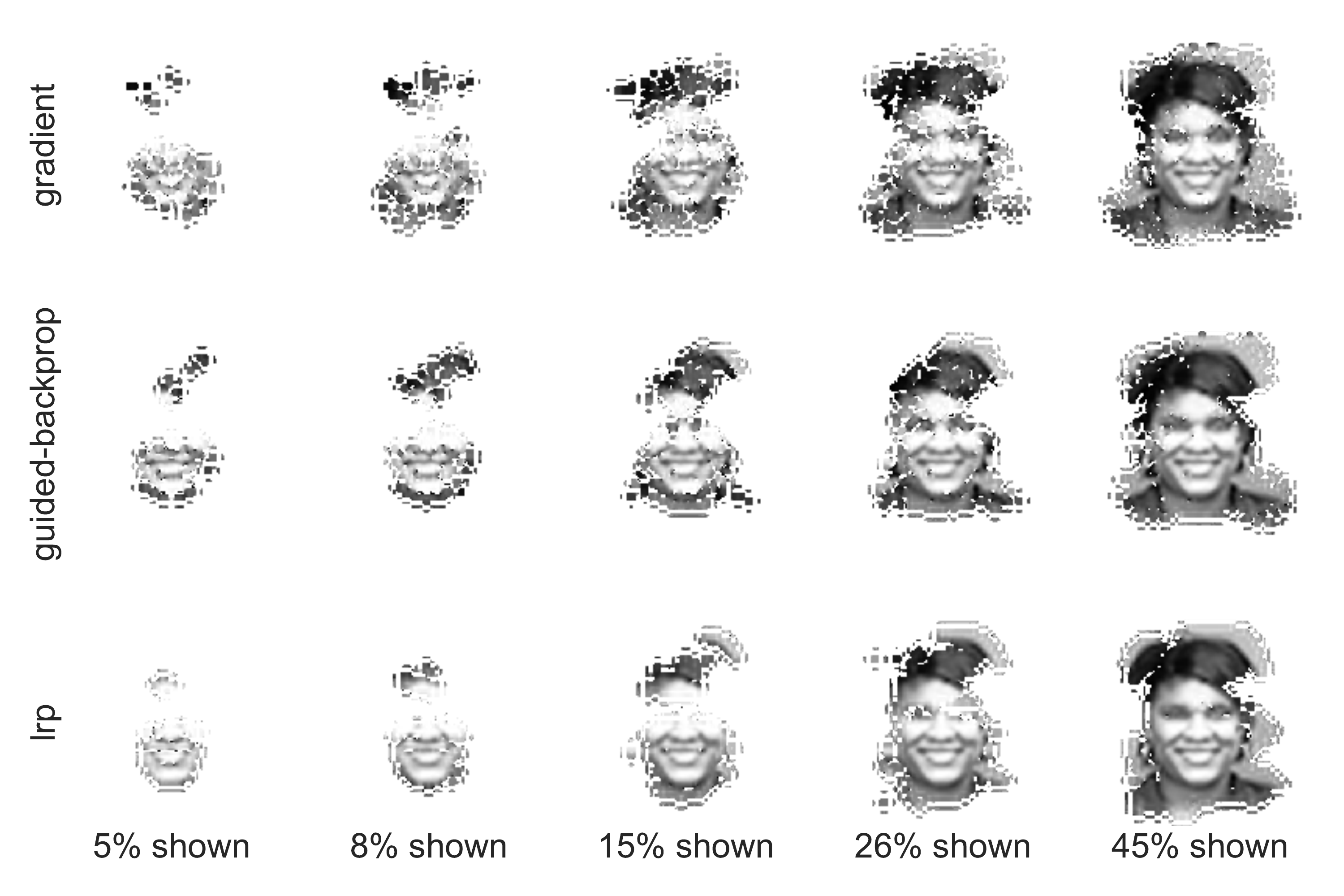}\\
%\includegraphics[width=\columnwidth]{figures/example_angry}
%\caption{Examples of masked images with happy emotional expression (\textit{top three rows}) and angry expression (\textit{bottom three rows}) for different interpretability methods (gradient, guided-backprop \cite{Springenberg2014} and LRP \cite{Lapuschkin2017}, shown in rows) and mask sizes (shown in columns)} 
\caption{Example of masked images with happy emotional expression for different interpretability methods (gradient, guided-backprop \cite{Springenberg2014} and LRP \cite{Lapuschkin2017}, shown in rows) and mask sizes (shown in columns)} 
\label{fig:example}
\end{figure}

\paragraph{Interpretability Methods}
We compared three different interpretability methods for the EmoPy computer vision model,  a) the gradient of output w.r.t. the input image, b) {\em Layerwise relevance propagation (lrp)} \cite{Lapuschkin2017} and c) {\em Guided backpropagation} \cite{Springenberg2014}. 
%
%%
%\begin{itemize}
%\item {\em Gradient}: The gradient of output w.r.t. the input image
%\item {\em Layerwise relevance propagation (lrp)} \cite{Lapuschkin2017}
%\item {\em Guided backpropagation} \cite{Springenberg2014}
%\end{itemize}
%%
For all methods we used the implementation in the \texttt{iNNvestigate!} package \cite{Alber2018}. All methods were used with their default hyperparameters. For the LRP approach we used the \texttt{sequential\_preset\_a} variant provided in the package. 
The list of methods is not meant to be exhaustive, but we chose these methods based on recommendations of the \texttt{iNNvestigate!} package. 
%The main purpose of this work is to illustrate that the combination of psychophysical methods and ML can be helpful for quantifying the usefulness of interpretability methods. For the sake of simplicity, we deliberately restricted the set of interpretability methods to just three methods that other experts in the field recommended to us as useful. 

\paragraph{Quality of Explanations}
%\label{sec:quality_metrics}
We compute two kinds of quality metric for each interpretability method, one with no humans in the loop (NHIL) and one approach based on psychophysical experiments with humans in the loop (HIL). In both settings we use all three interpretability methods to compute scores for each pixel in the image. We rank the pixels according to their score and mask a certain percentage of pixels. The percentages of shown pixels were ten logarithmically spaced values between 0 and 100 \cite{Fechner1860}. The masks showed 5,  6,  8, 11, 15, 19, 26, 34, 45, 60 percent of pixels of the image. Some example images for the emotional expression {\em happiness} are shown in \autoref{fig:example}. These thresholds were based on initial experiments with different thresholds in which we determined the minimum number of pixels needed to detect the emotion and the number of pixels needed to enable most subjects to correctly classify the image. 

\paragraph{Psychophysical human in the loop (HIL) metrics}
For each percentage of pixels shown we measured the annotation accuracy as well as reaction times in the emotional expression classification task. 

\paragraph{No humans in the loop (NHIL) metrics}
%When developing a new interpretability approach it is most convenient for researchers to iterate quickly on model improvements and to validate the improvements with tests that are ideally fast and can be conducted without humans in the loop. Most of these NHIL metrics perturb the input data in some way that takes into account the feature scores provided by an interpretability method. For instance in \cite{Samek2017} the authors replace small patches in an input image with noise and evaluate the predictive performance for each perturbation of the data. 
Following standard perturbation approaches \cite{Samek2017} we slightly modify the perturbations to match the conditions used in the psychophysics experiments. In particular we mask a certain percentage of pixels and feed the masked image to the convolutional neural network to obtain a prediction. To evaluate the interpretability quality we evaluate the predictive performance of the EmoPy model on masked images.
%
%\paragraph{Psychophysical human in the loop (HIL) metrics}
%In order to quantify the quality of interpretability methods in HIL psychophysical experiments we adopt the ideas from \cite{schmidt2019quantifying}, 
%%
%\begin{enumerate}
%\item Interpretability is associated with {\em intuitive understanding}
%\item Intuitive understanding leads to fast and accurate decisions
%\end{enumerate}
%%
%Accuracy and speed of AI-assisted decisions can give insights into the cognitive load inherent to understanding of ML predictions. When an explanation is intuitive we will follow it without too much thinking; but when we need more time to digest an explanation, its relative interpretability quality is lower compared to other explanations. More importantly, when ML assisted decisions are followed quickly even in cases when the ML predictions were wrong, this is a clear sign of unhealthy algorithmic bias. Evaluating both, reaction time and accuracy of annotations, can thus provide authentic and quantifiable metrics of interpretability quality. 
%Based on these ideas we measured the annotation accuracy as well as reaction times in the above emotional expression classification task. In the experiments we systematically controlled the amount of pixels unmasked by a given interpretability method to investigate the dependency of the signal strength and the interpretability. 

\paragraph{User Interface and Experimental Design}
We built the user interface using the open source library jsPsych \cite{deleeuw2015}. 
%The library provides basic features to design a psychological experiment running in the browser. In our case, we used the package to build an experiment timeline that showed the image stimulus with an html button below to capture the annotation provided by the experimental subjects.  
%
In each trial of an experiment we show the same image with increasing percentages of pixels shown and ask the subjects to report whether the expression was happy, angry or not certain. As we used ten different mask sizes from 5\% to 60\% of all pixels in the image, subjects saw a series of ten images, an example is shown in \autoref{fig:example}. At the last image, when 60\% of the image was shown, all subjects correctly identified the emotional expression, see also \autoref{fig:accuracy}. 
The entire experiment was designed to be completed in about 10 minutes, based on a pilot experiment. For each label five images were shown, which resulted in $5~\text{(images)}~\times 2~\text{(classes)}~\times 3 ~\text{(interpretability methods)} \times 10~\text{(mask sizes)} = 300$ images in total that were annotated by each subject. 
%The image stimulus is displayed in the format as shown in \autoref{fig:experimentinst}. 
For each subject the order of the trials was randomized, so each subject has seen each interpretability method and source image in a random order, but the order of unmasking the image was always the same. In total 62 subjects participated in the experiment.
%
%\begin{figure}
%\centering
%\includegraphics[width=.5\columnwidth]{figures/experiment-trial}
%\caption{User interface of a stimulus shown in the experiment. The image stimulus for a given mask size is located in the center of the page. Below the image buttons for annotating the image with the emotional expression \textit{anger} and \textit{happiness} are shown, along with an \textit{I don't know} option.} 
%\label{fig:experimentinst}
%\end{figure}
%
The experiments were conducted on the crowdsourcing platform Amazon Mechanical Turk. We payed all subjects the minimum wage in the country of the research institution of the authors, 11\$US per hour. 
%Mechanical Turk requires to show a preview of the experiment, before the worker accepts to participate. For the preview, we provided an instruction and an example trial.  
%In the main part of the experiment, after the workers agreed to participate, they will be first shown the number of trials they need to complete. When they proceed, the actual experiment will start and will be completed after the subjects have annotated 300 images. 
%!TEX root = psychophysics-XAI-4page.tex

\section{Results}
\label{sec:results}

In the following we compare the results of the psychophysical experiments with the results from the experiments without humans in the loop.

\begin{figure}
\centering
\includegraphics[width=6cm]{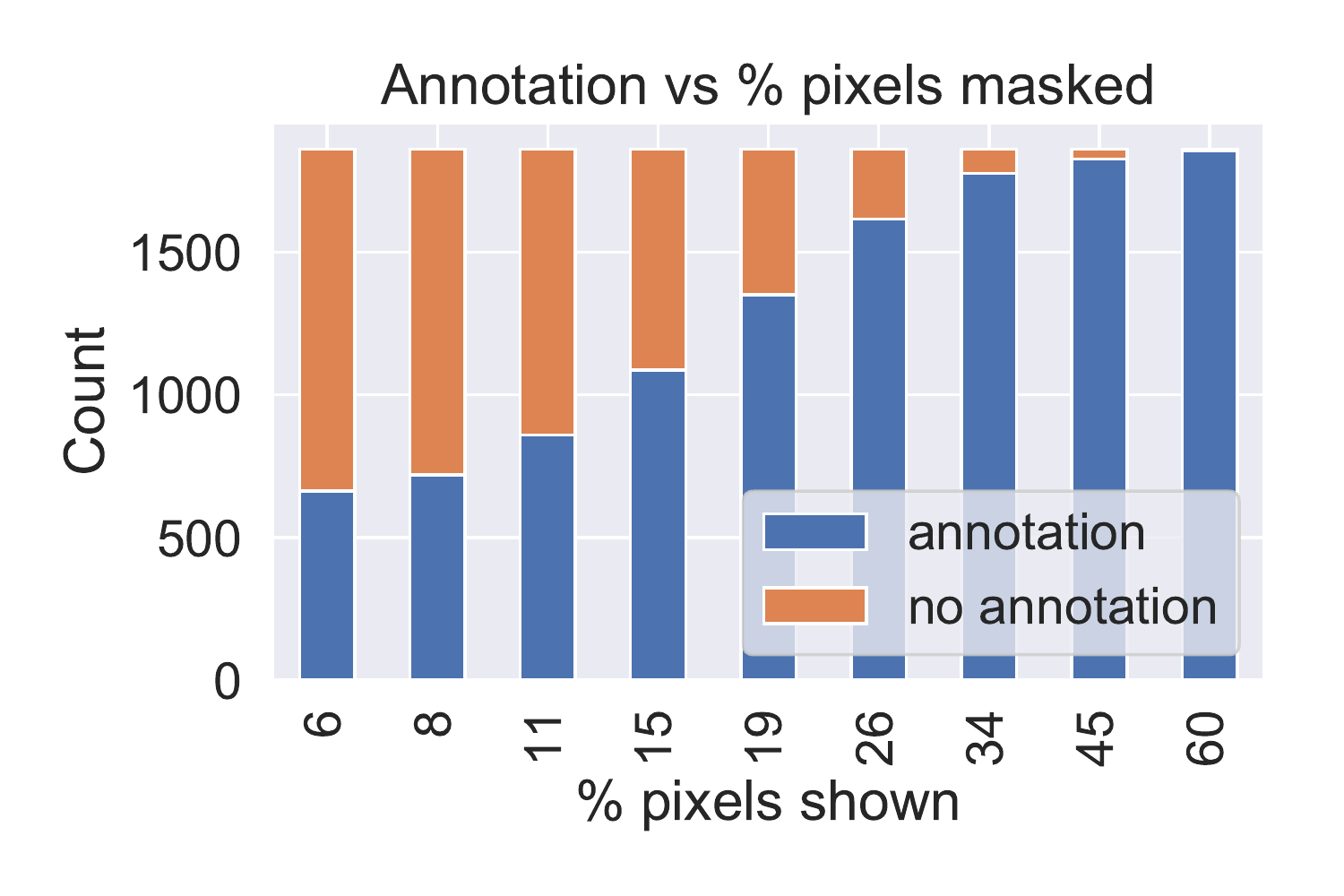}
\caption{Annotators' confidence, measured by counting how often they did not provide a label but the \textit{I don't know} label, as function of the mask size, aggregated over all interpretability methods. 
%When 6\% of all pixels were shown, 663 annotators detected an emotion and provided a label, while 1197 annotators did not detect an emotion and provided only the \textit{I don't know} label. 
When 60\% of pixels were shown, all annotators detected the emotion.}
\label{fig:uncertainty_all}
\end{figure}

%
%\paragraph{Psychophysical Experiments}
%\label{sec:psychophysics_results}

\paragraph{Annotators' uncertainty and interpretability}
We investigated the impact of each interpretability approach on the uncertainty of annotators by counting how often they did not provide an annotation but just the \textit{I don't know} label. Our results show a clear effect of the interpretability method as shown in \autoref{fig:uncertainty}. The simplest gradient approach leads to the highest annotator uncertainty and least number of annotations up to mask sizes of 45\% of all pixels. Guided BackProp \cite{Springenberg2014} in contrast leads consistently to the lowest annotator uncertainty and the highest number of annotations.
% Comparing Gradient and Guided BackProp we find that on average almost twice as many annotators are certain enough about their prediction that they provide an annotation when assisted with the Guided BackProp saliency map, compared to the Gradient explanation that more often led annotators to choose the \textit{I don't know} label. 
 This finding highlights the importance of quantitatively comparing transparency approaches. The extent to which human users of ML can profit from transparency strongly depends on the quality of the explanation provided. 

%\begin{figure}
%\includegraphics[width=8.5cm]{figures/rt_vs_threshold}
%\caption{Reaction times of all subjects for different interpretability methods and mask sizes. 
%Cognitive load related to processing the explanation appears to peak around mask sizes of 15\% pixels and decays for larger masks once annotators have detected the emotional expression shown.}
%\label{fig:rt}
%\end{figure}

%\paragraph{Reaction times reflect cognitive load of interpretatons}
%In \autoref{fig:rt} we show the reaction times for each experimental condition. 
%When most pixels are masked reaction times are low, as most subjects understand that they cannot make a correct prediction. For intermediate mask sizes around 15\% pixels shown, reaction times show a slight increase reflecting the increased cognitive load. For larger mask sizes, the reaction time decreases, as most subjects have provided an annotation already and keep clicking that label.

%
\begin{figure}
\centering
\includegraphics[width=7cm]{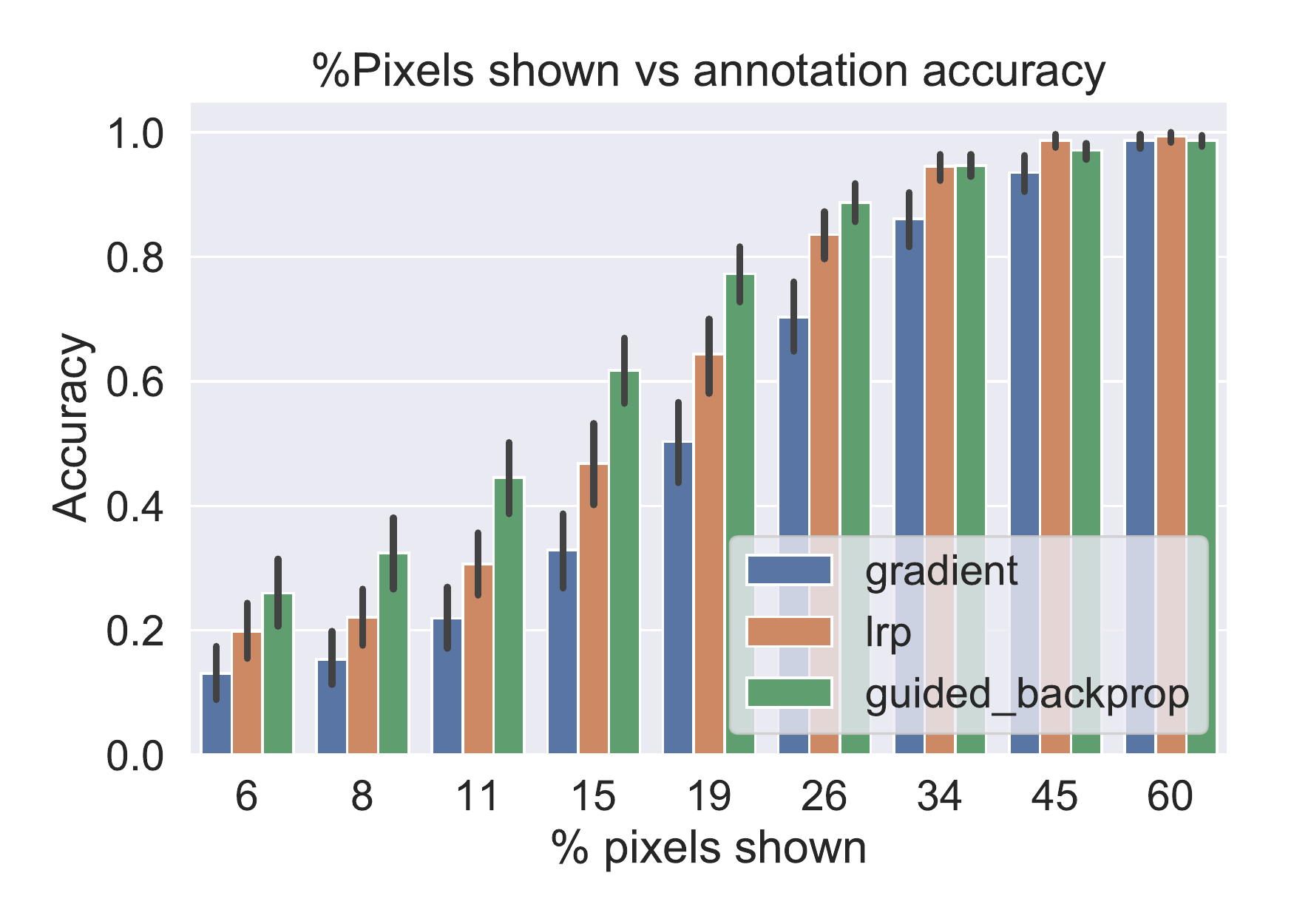}
\caption{Annotation accuracies of all subjects for different interpretability methods and mask sizes.% For small mask sizes, the annotation accuracy is as low as 20\%; 
when 40\% of the pixels are masked, all subjects reliably detect the correct emotional expression. 
For intermediate levels of masking, there is a clear ranking of interpretability methods, guided backprop achieves highest accuracies.}
\label{fig:accuracy}
\end{figure}

\begin{figure}
\centering
\includegraphics[width=6cm]{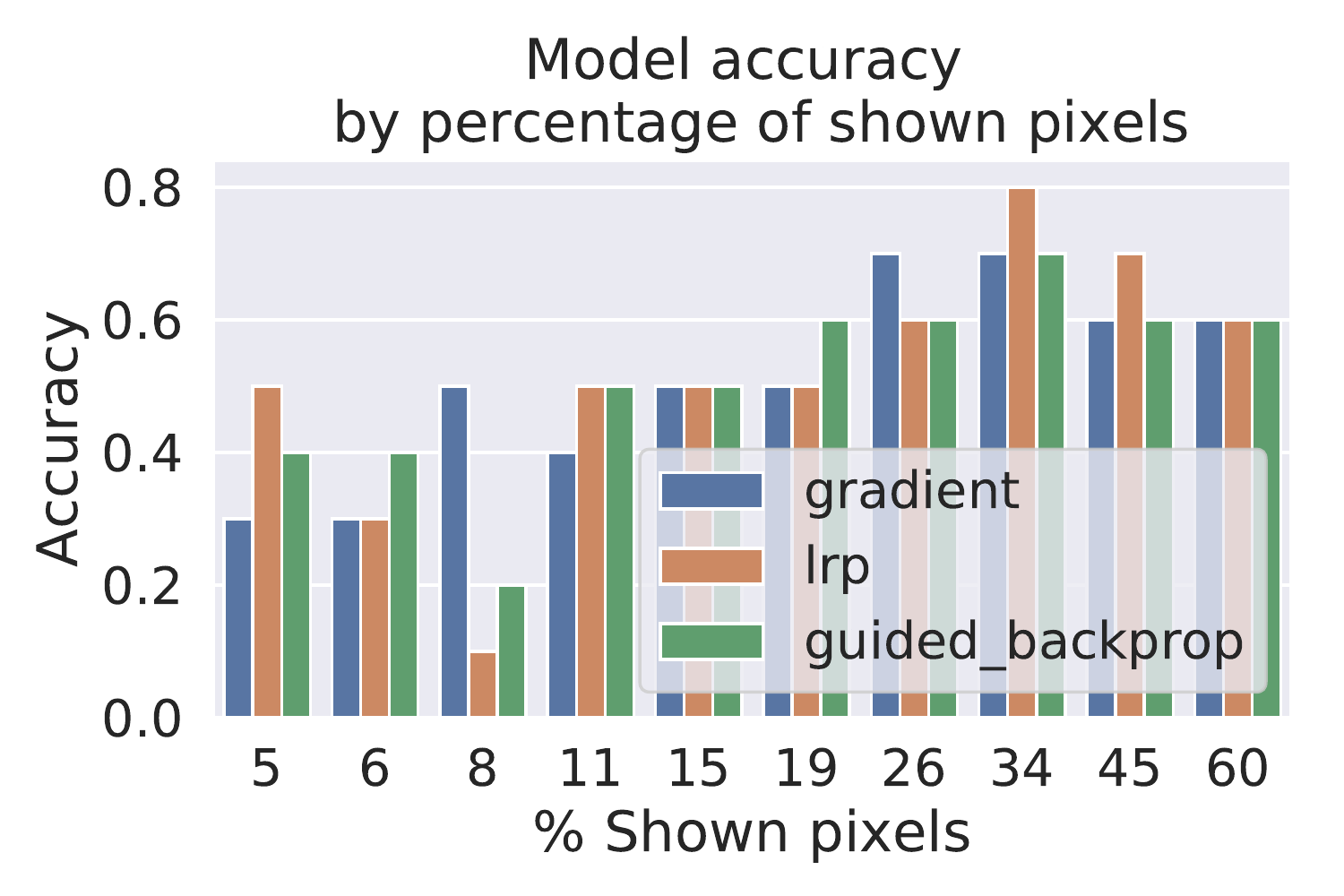}
\caption{Model prediction accuracies for different interpretability methods and mask sizes. In contrast to human annotators, there is no clear ranking of methods based on the accuracy of the ML predictions.}
\label{fig:accuracy_model}
\end{figure}

\begin{figure*}
\includegraphics[width=4.9cm]{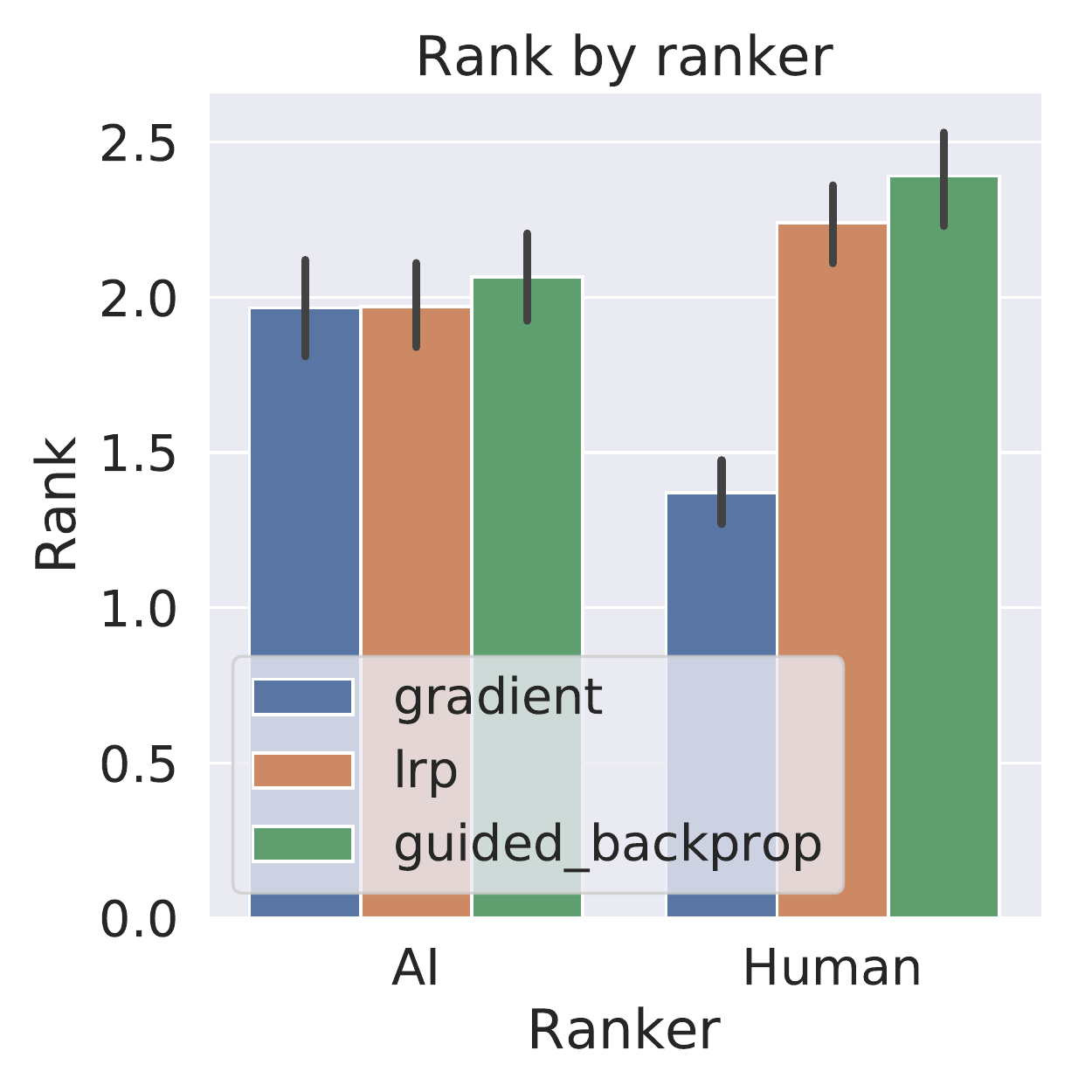}
\includegraphics[width=12.4cm]{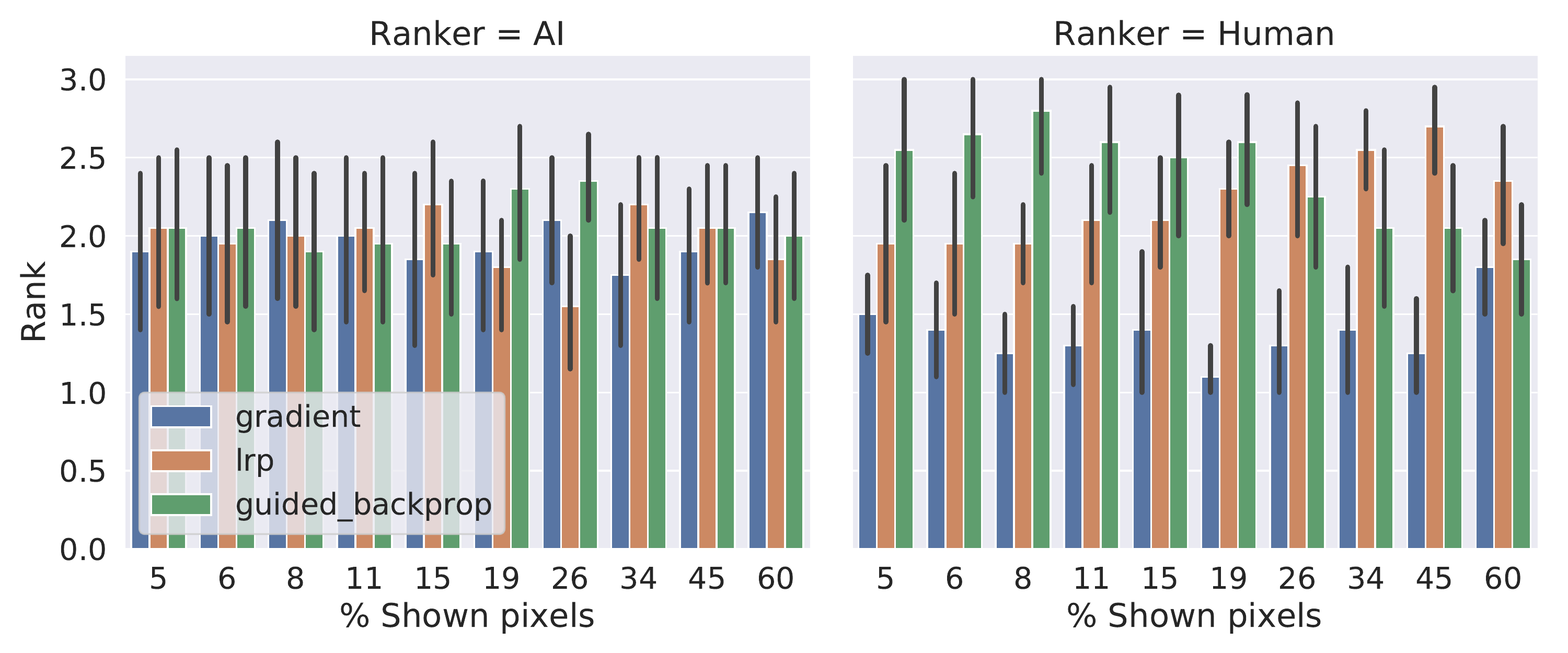}
\caption{XAI method quality ranked by annotation accuracy (ranker=human) or cross-entropy loss per image (ranker=AI). \textit{Left panel}: automated XAI rankings are instable while human rankings show robust pattern, Guided BackProp explanations are consistently best. \textit{Middle and right panel}: Ranks for each interpretability method computed on AI predictions and human annotators for each mask size. For most mask sizes human annotators' accuracy was significantly higher for the Guided BackProp approach, there is no clear winner for the AI interpretability quality metric.}
\label{fig:rank_vs_threshold}
\end{figure*}

\paragraph{Annotation accuracy distinguishes XAI methods}
%The most important metric for our purposes is the annotation accuracy for different interpretability methods. Higher quality explanations should lead to higher annotation accuracy. Indeed w
%We find that annotation accuracy clearly distinguishes the three interpretability methods used in our experiments. 
In \autoref{fig:accuracy} we show the annotation accuracy, averaged across subjects, for increasing mask sizes and all three different transparency approaches. Explanations using the plain gradient approach consistently led to the lowest annotation accuracy. The layerwise relevance propagation approach (LRP) \cite{Lapuschkin2017} yielded slightly better annotation accuracies and annotators assisted with the Guided BackProp explanations \cite{Springenberg2014} were consistently better than all other annotators. This effect was strongly dependent on the mask size and most pronounced for intermediate mask sizes around 15\% of pixels shown. When more than 45\% pixels were shown, subjects could detect the emotional expression reliably in all conditions. These results suggest that psychophysical experiments can serve as a robust quality indicator for interpretability methods. 

%
%\subsection{Metrics with no humans in the loop}
%\label{sec:model_results}

\paragraph{Instability of automated XAI metrics}
%Next to the human in the loop experiments we also performed more standard experiments in which we tested the three interpretability approaches under the same experimental conditions as in the psychophysical experiments. For each mask size and interpretability method we computed the predictions of the EmoPy \cite{emopy} model and computed the accuracy across all images for each condition. The results are shown
%
We show the results of the automated XAI quality metrics without humans in the loop in \autoref{fig:accuracy_model}. 
%At around 30\% of all pixels the model achieves the highest performance, that is better than the optimal performance on the test. When masking more pixels the prediction accuracy decreases irregularly without any specific trend, unlike in the case of the human annotators. 
The most notable difference to the human in the loop experiments is that the across all mask sizes there is no clear winner and in some cases the method that scored worst in the psychophysical experiments, Gradient, achieves the best accuracies when evaluating it on ML model predictions alone. 

%Less important but interesting is also that while the general trend of lower accuracies with smaller masks is the same for both humans and machines in our experiments, human cognition tends to have a different sensitivity to the amount of pixels masked. While humans achieve a lower performance than the ML model when only 6\% of pixels are shown, this effect is reversed when more than 45\% of pixels are shown: annotators do not make mistakes in the emotional expression classification task  and the ML model only achieves accuracies around 60\% in those conditions. 

%Both of these findings demonstrate that human cognition and machine cognition share some properties but are very different in others. In particular interpretability metrics that are purely based on machine predictions do not seem to capture what makes an explanation useful for humans. 
%

\paragraph{Comparing NHIL and HIL metrics}
%\label{sec:comparisons}
%While the previous sections focussed on each metric individually we also compared the metrics obtained in psychophysical experiments with humans in the loop, see \autoref{sec:psychophysics_results}, with the metrics obtained by conventional offline no human in the loop (NHIL) approaches, see \autoref{sec:model_results}. 
For the comparison between XAI metrics with and without humans in the loop we grouped the data by image, mask size and ranked each interpretability method by the average accuracy across all subjects for each image (human in the loop metric) and the cross-entropy loss incurred by a prediction for each interpretability method (no-human-in-the-loop metric). The resulting rankings are shown in \autoref{fig:rank_vs_threshold}. In the left panel the aggregated ranks across all mask sizes show that despite the transformation of the metrics into ranks, the two types of metrics are not very similar. Interpretability metrics obtained in the psychophysical experiments show a clear and robust ranking, while the rankings obtained by ML predictions do not allow to distinguish the three methods in terms of their interpretability. This is also reflected in the two right panels, which show the same data as in \autoref{fig:rank_vs_threshold}(\textit{left}), but split into all mask size conditions. The average ranks of the psychophysical experiments show the same clear pattern as the aggregate metrics, Guided BackProp is better than LRP which is in turn better than the plain Gradient explanation. In contrast it is difficult to single out the best interpretable explanation based on the machine based NHIL metrics; there is no significant difference between the methods for most thresholds. 
\section{Conclusion}
\label{sec:conclusion}
%
%Methods that increase transparency of ML systems have become a major focus of research. 
%Despite substantial advancements in the field and a plethora of methods available for rendering ML model predictions more interpretable, there appears to be no gold standard evaluation method for interpretability quality \cite{Guidotti2018}. 
Reliable and quantitative measures for evaluating interpretability are however a fundamental prerequisite for designing and improving transparent ML systems \cite{Adebayo2020}. 
Often interpretability evaluations without humans in the loop \cite{Samek2017, Adebayo2018} cannot be directly compared with humans in the loop metrics \cite{Hase2020}. 
%Both approaches usually do not follow the same experimental design which makes comparisons across studies difficult. 
%To the best of our knowledge there are few studies that use the same experimental conditions for humans and machines when evaluating interpretability methods and that relate results from human in the loop experiments to evaluations without humans. 
%
In this study we used psychophysical experiments to directly compare XAI quality metrics with and without humans in the loop. 
%
%Our results demonstrate that while psychophysical experiments allow to derive robust and clear rankings of interpretability quality, interpretability metrics obtained with ML predictions alone do not show a clear ranking of interpretability methods. More importantly our results also show that the metrics computed without humans in the loop are not only instable, they are also not representative of the rankings obtained in psychophysical experiments. 

Our findings demonstrate that not only do machine based interpretability metrics not allow for a clear comparison of interpretability methods, more importantly these metrics are not representative of what is relevant for interpretability by humans either. 

These results highlight the potential of standardized psychophysical tests for the evaluation of ML methods and indicate that evaluations of interpretability should not rely exlusively on experiments without humans in the loop.

\bibliography{references}
\bibliographystyle{icml2021}

\end{document}